    \documentclass[%
     aip,
    rsi,%
     amsmath,amssymb,
     reprint,%
    ]{revtex4-1}
    
    \usepackage{graphicx}
    \usepackage{dcolumn}
    \usepackage{bm}
   \usepackage[colorlinks = true,
            linkcolor = blue,
            urlcolor  = blue,
            citecolor = blue,
            anchorcolor = blue]{hyperref}

\begin{document}
    
    \title[Branched flow]{Branched Flow}
    
    \author{Eric J Heller}
    \homepage{https://www-heller.harvard.edu}
     \affiliation{Harvard University
     Department of Physics and Department of Chemistry and Chemical Biology}
     \author{Ragnar Fleischmann}
    \homepage{https://www.ds.mpg.de/meso}
    \affiliation{%
    Max Planck Institute for Dynamics and Self-Organization (MPIDS), G{\"o}ttingen, Germany
    }%
    \author{Tobias Kramer}%
    \homepage{https://www.zib.de/members/kramer}
    \affiliation{ 
    Zuse Institute Berlin (ZIB), Berlin, Germany
    }%

    \date{\today}
    \begin{abstract}
    
    {\large
    In many physical situations involving diverse length scales, waves or rays representing them travel through media characterized by  spatially  smooth,    random, modest refactive index  variations.
       ``Primary'' diffraction (by individual sub-wavelength features)  is absent. Eventually the weak refraction    leads to  imperfect   focal ``cusps''. Much later,  a  statistical regime  characterized by momentum diffusion is manifested.    An  important intermediate regime is often overlooked, one that is   diffusive  only in an ensemble sense.  Each realization of the ensemble  possesses dramatic  ray limit structure that guides the waves (in the same sense that ray optics is used to design lens systems).  This structure is  a universal phenomenon  called branched flow. Many important phenomena develop in this intermediate regime.       Here we  give  examples and some of the physics of this emerging field.

    }
    
    \end{abstract}
    \maketitle
        \begin{figure*}
    	\includegraphics[width=6in]{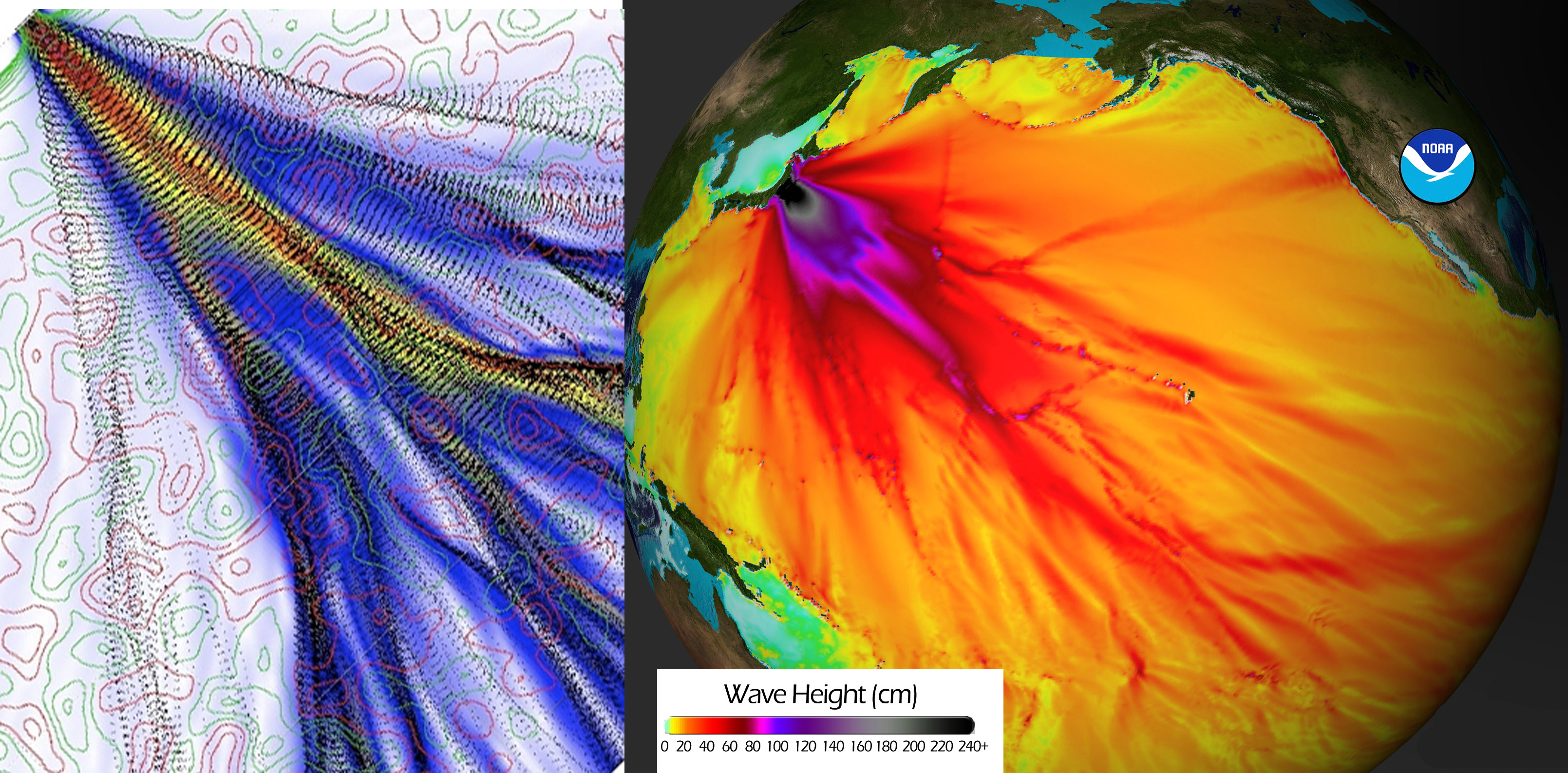}
    	\caption{ \textbf{Universality: Semiconductor electron flow and Tsunami wave propagation}  (Left) Micron scale ray (small black dots) and wave (color) propagation from a small quantum point contact source over a random potential field modeling semiconductor electron flow. (Right)  A reconstruction of the March 2011 tsunami in Pacific ocean by the  NOAA Center for Tsunami Research shows pronounced height fluctuations of the tsunami forming a branched pattern. [Copyright:National Oceanic and Atmospheric Administration (NOAA) / PMEL / Center for Tsunami Research] 
    	\label{fig:tsunamis}}
    \end{figure*}

    \begin{quotation}
    Electron waves refracted by  disorder in high mobility semiconductors, ocean waves deflected by surface eddies,  sound waves refracted in the turbulent atmosphere or refracted underwater by temperature, salinity, and pressure variations,  tsunami waves refracted by variations in ocean bottom depths, gravitational lensing of light   by galactic clusters including associated dark matter, deflection of light in media with refractive index fluctuations, large scale structure of the universe, light refracted by living tissue, and   the earliest studied example we know of: pulsar-generated microwaves refracted by ionized interstellar clouds.   These are some of the examples of waves (or rays in the classical limit) in their natural environment  that are weakly deflected (forward scattered) by  essentially random fluctuations in a medium with  correlation lengths longer than a wavelength. These systems exhibit a long neglected but important transport regime, transitioning between ballistic and  diffusive behavior. 
    \end{quotation}
    
\vskip .1in
\noindent{\bf Introduction: tsunami waves in shallow water}
\vskip .1in

     What is certainly the most fearful example of branched flow is  a good platform to introduce the phenomenon: ocean tsunami waves. Subsurface earthquakes  (or large coastal landslides) may excite energetic surface gravity ocean waves of extremely long wavelengths (in the range of several tens to   hundreds of kilometers). The waves, traveling  hundreds of kilometers an hour, are propagating in very shallow water relative to their wavelength.       A March 2011  magnitude 9 earthquake offshore of Tohoku Japan  initially caused waves to propagate outward like  ripples from a rock thrown into a shallow pond.  However,  for those waves heading away from shore, weak refraction  caused by variations in bottom depth accumulate to create dramatic  branch-like energy or flux density variations, as illustrated in the NOAA tsunami wave energy plot, Fig.~\ref{fig:tsunamis}. This image is typical of  branched flow.  See the NOAA website \href{https://nctr.pmel.noaa.gov/honshu20110311/}{https://nctr.pmel.noaa.gov/honshu20110311/} for other images including a revealing animation. 
     
    The  speed of the  tsunami wave is proportional to the square-root of the ocean depth, averaged on scales of the order of the wavelength. Underwater mounds  act as focusing lenses~\cite{Berry2007}, depressions act  as diffusing lenses.     Even depth fluctuations  of only a few percent can lead to formation of branches.  It was shown recently that the bottom depths of earth's oceans are not known well enough to accurately predict the locations of   distant tsunami branches\cite{Degueldre2016}.  Life saving wave energy predictions  could be done well before destructive energy reached  shore, if the ocean depths were better known.

\begin{figure*}
    	\includegraphics[width=5.5in]{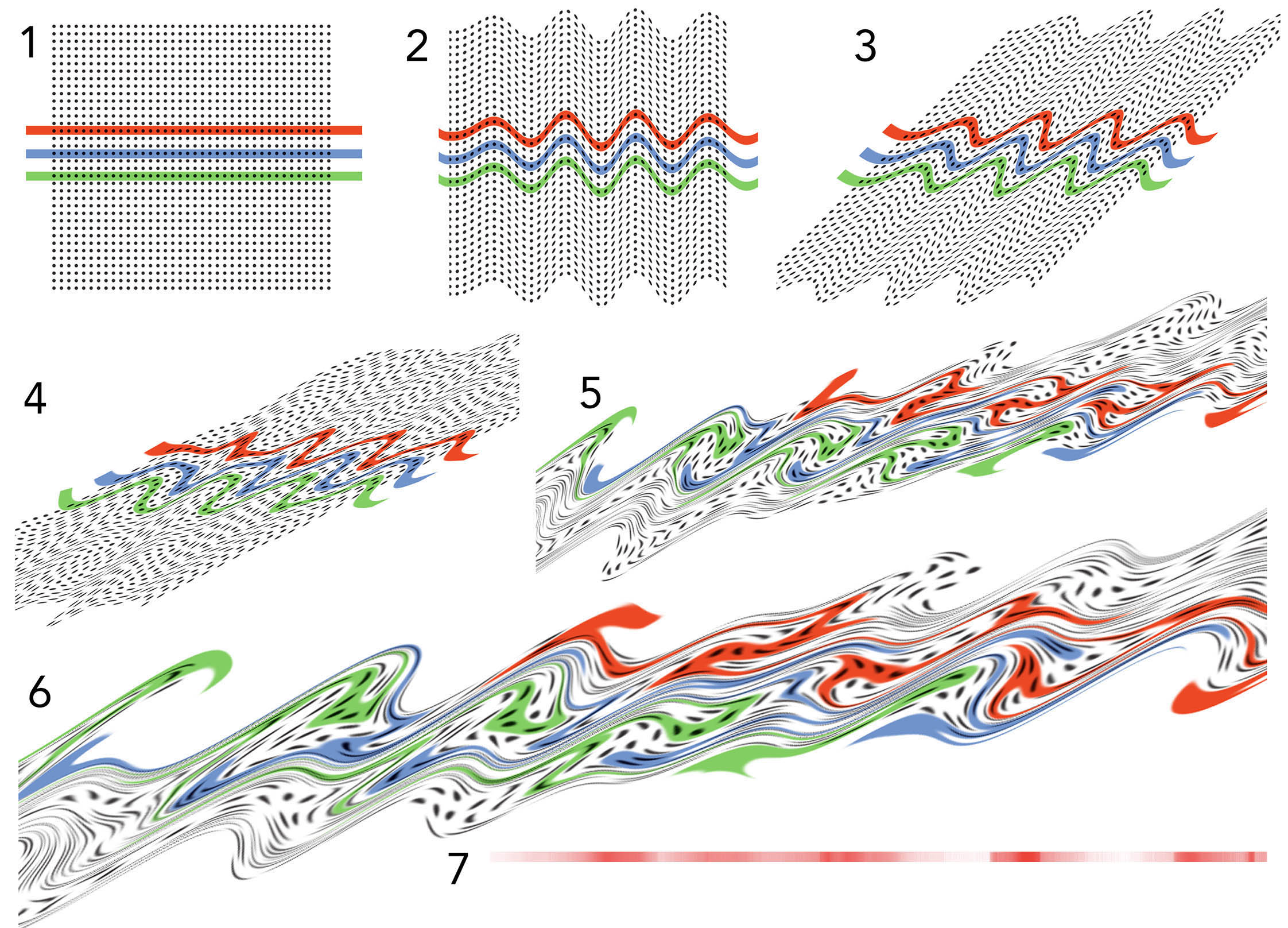}
    	\caption{ \textbf{Classical kick-drift (thin lens) model implemented in Photoshop} Frame 1 shows the phase plane decorated with circles and bands that will reveal the effects of the map. The dynamics is described in the text. Note some of the original circles are  greatly stretched and distorted, while others are nearly unchanged, indicating stable or nearly stable zones, surviving here for six kick-drift iterations. The formation of branches corresponds to the color accumulations resulting from the vertical projection of the phase space onto coordinate space. The bar on the lower right (frame 7) records the density of the distorted red band.  The three initial colored bands correspond to different ranges of initial propagation directions, each a ``fuzzy manifold''.
    	\label{fig:kd}}
    \end{figure*}

    Branched flow involves many small angle refraction events caused by smooth weak fluctuations in the medium, each spanning several wavelengths or more (for tsunamis these are ocean bottom depth variations). Classical ray tracing is applicable, depending on the wavelength relative to the size of the refractive features.  The  source   of the waves needs to be localized in some way that is best described as a surface or ``manifold''  in phase space, such as emission  in many directions from a spatially localized point source (as from the Tohoku earthquake or a pulsar), or a spatially extended source with initially restricted propagation directions, like a plane wave. Many real  sources  correspond to  ``fuzzy'' manifolds, only semi-localized in phase space;  physical examples are light rays from the disk of the sun  or waves leaving a storm from a region on the ocean,  with a range of  propagation directions. The branching phenomenon is still rich and dramatic for these ``averaged'' sources. The colored band in figure~\ref{fig:kd} below are fuzzy manifolds.

    \vskip .1in
    \noindent{\bf Ray modeling}
    \vskip .1in

    Although the gross branched structure follows from classical ray tracing,  the signal is certainly decorated with interference effects for coherent sources like pulsars (see below). At long propagation distances, the differences between ray simulations and the actual wave evolution can become profound.  Not nearly enough is known about the wave-ray correspondence breakdown. 
    
\begin{figure*}
    	\includegraphics[width=6.8in]{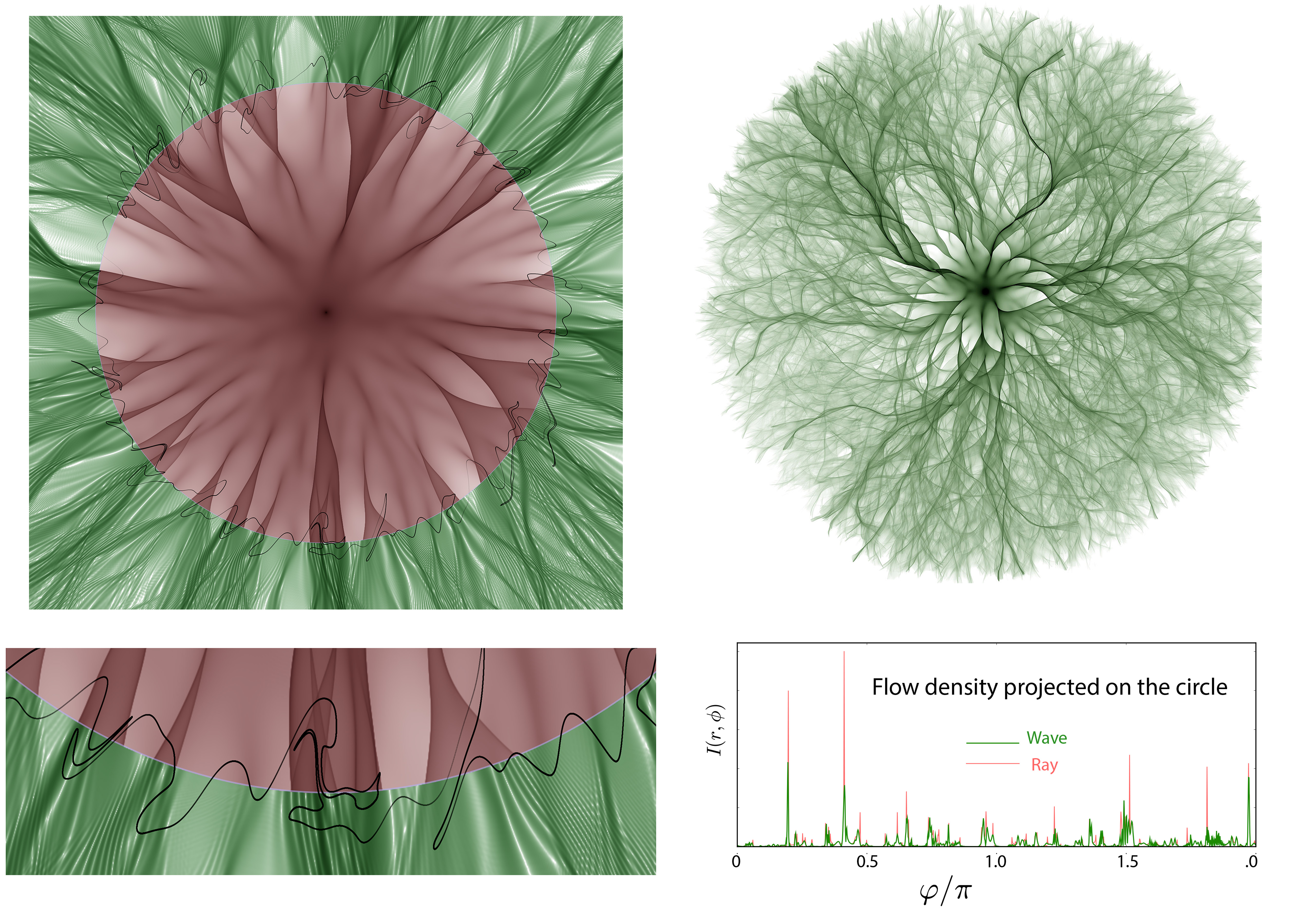}
    	\caption{ \textbf{Classical and quantum branched flow} On the top left, the pink zone shows a classical ray tracing simulation, while the outer green zone depicts a wave equation simulation, for a point source in the center of the image sending flux of speed momentum uniformly in all directions, over the same weak random potential (not shown). Below left, the flow is analyzed in more detail, with the black line giving the ray momentum parallel to the circle as a function of position on the circle. At  bottom right the remarkably nonuniform classical ray (red) and wave (green) flow density is depicted as it intersects the pink-green radius.   On the top right, a purely ray tracing simulation was followed for more generations of  fold catastrophies. It possesses striking stable branches, which result in strong, thin flux density visible as thin, curving lines. 
    	\label{fig:colorbranch}}
    \end{figure*}

     A picture worth  a thousand words is seen in figure~\ref{fig:colorbranch}, which abruptly changes from ray to wave solutions for 2D transport over a smoothly varying weak random potential  (not shown). There is a point source of the waves or rays at the center.   Obvious agreements and some subtle differences  between the ray tracing   (pink) and the wave solutions are seen. 
    
    There is a very efficient way to summarize the detailed classical flow, by erecting a surface (in this case a circle) perpendicular to the average flow, and plotting the position (here the angle $\theta$) of intersection on the surface versus the momentum component $p_\theta$ along the surface, for each ray that penetrates the surface (a \emph{Poincar\'e surface of section}). Circles of different radii reveal the evolution of the flow in phase space.  The evolution is equivalent to an area preserving nonlinear map of the phase plane into itself, as seen in figure~\ref{fig:colorbranch}.   
    \vskip .1in
    \noindent{\bf Photoshop dynamics}
     \vskip .1in
     It is very instructive to define a discrete, nonlinear,  ``kicked'' map for rays. This closely mimics following rays  passing through a succession of independent thin lenses or phase screens.  Each iteration of the  map applies a new   random (with some correlation length)  impulsive momentum kick; the momentum is the component normal to the mean flow.  The  kick is followed by a  shear corresponding to free drift,  defining a  single kick-drift cycle.  The kick-drift model is equivalent to the paraxial limit of a series of thin lenses or phase screens in the optical case. In a N-dimensional kick-drift model the $(N+1)$st coordinate is time, whereas it is a physical length in an $(N+1)$ dimensional coordinate space for a series of thin lenses.
     
     The kick-drift model is easy to define as  a point-to-point area preserving mapping of the phase plane into itself, under 
        \begin{align} 
                  p_{n+1} &= p_n -\frac{dV_n(x)}{dx}\Bigg |_{x=x_n}\mathrm{(kick \ step)} \nonumber \\
          q_{n+1}&= q_n +  p_{n+1} \ \ \mathrm{(drift \ step)} 
        \end{align}
The potential $V_n(x)$ changes randomly with each $n$, however retaining certain statistical properties such as a correlation length in $x$.
      
      In the plane this protocol can be implemented by standard image processing using software like Photoshop or the python script provided at \href{https://github.com/tobiaskramer/branch}{https://github.com/tobiaskramer/branch}.
      
      Starting with some marked  zones in the image representing phase space (for example using lines and small disks as in figure~\ref{fig:kd}), a random (with a given correlation length) spatially varying vertical   momentum shear is applied (frame 2   in figure~\ref{fig:kd}).
    Frame 3 depicts the subsequent effect of   free drifting,linear shearing to the right with positive momentum and to the left with negative,  while frames 4 through 6 show the result of further random kick-drift cycles. Finally frame 7 shows the projection of the red band region of the first frame onto coordinate space.  The    branches are  revealed as brighter red zones in the projection;  note that most zones have more than one red region in the phase plane contributing to them.  In the wave or quantum version, these separate contributions would carry their own phase and amplitude, interfering constructively or destructively. The qualitative features seen here apply even if thousands of kick-drift cycles are involved, with much smaller kicks.
    
     The fate of fuzzy manifolds can be seen in figure~\ref{fig:kd}, where the colored bands span a small range of initial momenta and are therefore fuzzy manifolds. See also the discussion surrounding  figure~\ref{fig:freak} below. 
    
    Some stable or nearly stable zones survive the six kick-drift cycles, as seen by the black disks that are still relatively undistorted at step 6

\begin{figure*}
    	\includegraphics[width=6in]{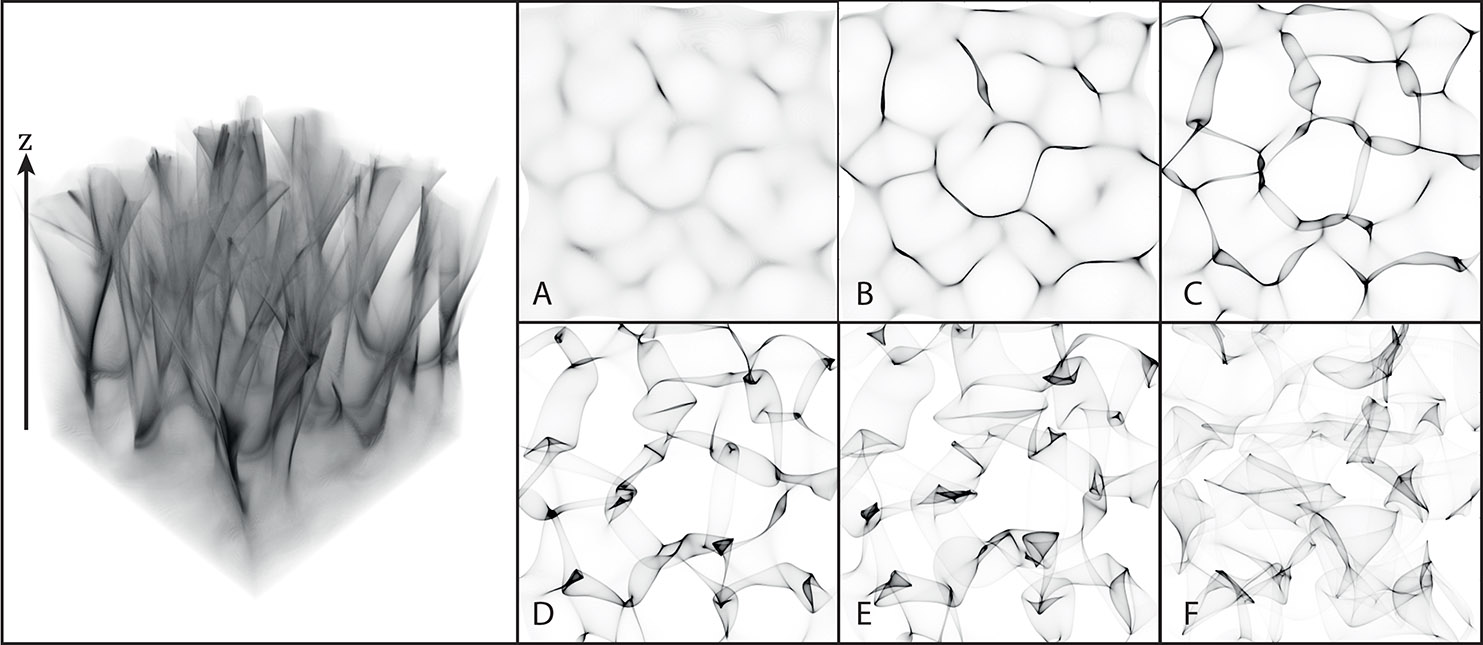}
    	\caption{ \textbf{Three dimensional branched flow} (a) Classical ray tracing showing three dimensional branching  resulting from propagation through a medium with modest random correlated refractive index changes. The flow began at the base, uniform in density with every ray initially heading vertically. At the right are seen two dimensional slices through the three dimensional flow, showing accumulation of the flow into strong tubes or branches.
    		\label{fig:branch3d}}
    \end{figure*}

     \vskip .1in
    \noindent{\bf Dimensionality}
    \vskip .1in
    The branching phenomenon applies in three as well as two dimensional wave flow. Figure~\ref{fig:branch3d} is a ray tracing simulation of a three dimensional propagation, such as might occur with sound waves traveling through an atmosphere with spatial temperature and velocity fluctuations   inducing  refraction.  Nearly everyone has noticed  irregular    loudness variations of the sound of  a jet far overhead.  The loudness pattern on the ground is a moving (as the source moves) variation like that seen with light on a pool bottom; see figure~\ref{fig:branch3d}.  Indeed, the chance focusing of damaging sonic booms  onto sensitive places like surgery rooms contributed to the 1971 banning of commercial supersonic flight over land in the United States. 
    \vskip .1in
    \noindent{\bf Characteristic length scales, features, diffusion}
    \vskip .1in
    	One can ask of figure~\ref{fig:colorbranch}: what determined the average angular distance $d$ between the first focal cusps, and what determines the radial distance $L$ from the origin to the typical first cusp~\cite{Kulkarny1982,Kaplan2002}? Suppose the flow is classical with trajectories having  energy $E$, but the smooth potential undulations are typically of height or depth $\epsilon << E$.  Then simple arguments about the deflection of rays by the undulations show that the distance between cusps is given by the correlation length $d$ of the potential bumps, and the radial distance $L$ to the first focal cusps is given by 
        $$L  = d \left (E\over \epsilon\right )^{2/3}.$$
      This and other statistical properties of the branched flow are widely independent of the details of the random potential, i.e.~the functional form of the correlations, and are thus in the same sense \emph{universal} as the mean free path~\cite{Metzger2010,Metzger2014}. Over an ensemble of random potentials, the direction of momentum initially delta peaked at $\theta = 0$ diffuses as 
       $$P(\theta,t) = \frac{1}{ \sqrt{4 \pi D t} }e^{-\frac{\theta^2}{4 D t}} $$
      with $D\propto \epsilon^2 /E^{3/2} d$.
    The initial direction of travel decays as 
    $$\int P(\theta,t) \cos(\theta) d\theta = e^{-t/\tau}$$
    where   $\tau = 1/D  $. Thus in weakly refracting media, $\epsilon/E\ll 1$, the mean free path  $\ell=\left<\left|v\right|\right>\tau\propto\left(\epsilon/E\right)^{2}d$ is much larger than the typical branching length scale, i.e.\ $\ell\ll L$. Therefore there is a wide regime where wave propagation is dominated by branched flow. 
    The caveat of these kind of arguments is that they are ensemble results; individual cases (as in figures \ref{fig:colorbranch} and \ref{fig:tsunamis}) are in many aspects more revealing.
   
    The geometry of classical ray tracing branched flow,  wave propagation branched flow, and the relation between them   require more study. One aspect always mentioned is the caustics, i.e. fold catastrophies of the phase space flow projected onto coordinate space. However these alone do not account for the formation of strong branches.  Another measure introduced some years ago is the rarefaction exponent\cite{heller2003branching}, a measure of the stretching or compression of the initial manifold along itself, i.e. along the direction tangent to the manifold surface. Compression along this direction can pile up flux density to create branches without caustics or enhance branches with them.

    \vskip .1in
    \noindent{\bf Stable Branches}
    \vskip .1in
    
    Investigating sound propagation under the ocean, Wolfson and Tomsovic\cite{TOMSOVIC} first recognized the subclass of branches that can be termed ``stable'': Zones of initial conditions that, by accident, lead to stable or nearly stable evolution of nearby trajectories for some finite distance or time. 
    For some subclass of initial conditions it will happen that dilating and contracting directions of phase space successively almost cancel each other.  Clearly, the percentage of such ``lucky'' zones of initial conditions must decrease exponentially with time, but several nearly stable zone are identified as nearly undistorted disks surviving after six violent kick-drift episodes in figure~\ref{fig:kd}; stable or nearly stable branches can easily be identified in figure~\ref{fig:colorbranch}, top right.
    \vskip .1in
    \noindent{\bf Large scale structure of the universe; comets and pool bottoms: single kicks}
    \vskip .1in
    
    The popular Zeldovitch model of the large scale structure of the universe   results from a single early universe kick and drift episode, followed by ``sticky'' gravitational effects after caustics form.   It is thought that the  relatively uniform early universe  matter distribution possessed random fluctuations in velocities, down to some cut-off length scale. Or, the initial dynamics could be a slow gravitational acceleration  due to  nascent mass density fluctuations. Both mass and velocity variances are attributed to quantum fluctuations in the first instants of the universe. In both (or combined) cases there is a long period of free (or weakly accelerated) drift  of relatively tenuous and mostly non-interacting  matter.  This kick and drift cycle   leads to formation of caustics looking very much like those  in figure~\ref{fig:branch3d}, B, before gravitational effects changes the dynamics and things become ``sticky,'' halting the free drift. Following up on this idea with numerical simulations gives rise to structures very like the observed cosmic web of matter, including the large voids \cite{zeldovichApprox} and highlighted by J.~Peebles in his book on the large scale structure of the universe in the chapter ``caustics and pancakes'' \cite{Peebles1980}.
    
    This is single kick 3D branched flow, with time being the 4th dimension, i.e. the  flow coordinate. The initial kick-drift episode is entirely analogous to sunlight falling on a pool bottom after being refracted by  surface waves. Interestingly, the departure of matter from the undulating surface of a comet is much the same story, with matter being ejected everywhere predominantly normal to the local tangent plane\cite{comet}.

    \vskip .1in
    \noindent{\bf Pulsar microwaves}
    \vskip .1in
    
      Microwaves launched towards earth by pulsars are often refracted by  one or more intervening partially ionized interstellar clouds.  The waves   escape the clouds to begin  long free space traverses,  often followed by a second, third, ... cloud followed again by a free propagation episode,  finally reaching earth. The microwave source is an observational dream:  pulsed, broadband, and a virtual pinpoint in the sky.  The ionized clouds are dispersive (with propagation speed going as $1/\omega^2$) with total time delays measuring the total column density of electrons. It may take several minutes or more for the microwaves from a single pulse to arrive at earth, shortest wavelengths first.  The   refractive index variations within a cloud are surely multiscale.   The stunning microwave data   at a radio telescope like Arecebo  is taken as  a ``spectrogram", i.e. signal strength as a function of both time (typically on the order of hours) and frequency (typically displayed over a range of 10 MHz starting between 30 and 1500 MHz), well within uncertainty principle limits. The signal, traveling for many years through clouds and our mostly empty galaxy,  usually   varies dramatically by the minute, hour, day,... at a single telescope.  This is  due mainly to  ``scanning,'' i.e.  proper  motion of a pulsar moving relative to a cloud,  the earth presumably passing through different branches, and the interference of coherent microwaves that have taken different paths to reach the telescope.  
      
      Pulsar microwaves are the earliest example we are aware of (Pidwerbetsky PhD thesis, Cornell, 1988 \cite{Pidwerbetsky1988}) where ray tracing led to a clear depiction of the branched flow regime and some of its implications.  The seminal pulsar microwave paper by Cordes, Pidwerbetsky, and Lovelace\cite{osti_6398373} contains nuggets of Pidwerbetsky's thesis.  
       \begin{figure*}
    	\includegraphics[width=6in]{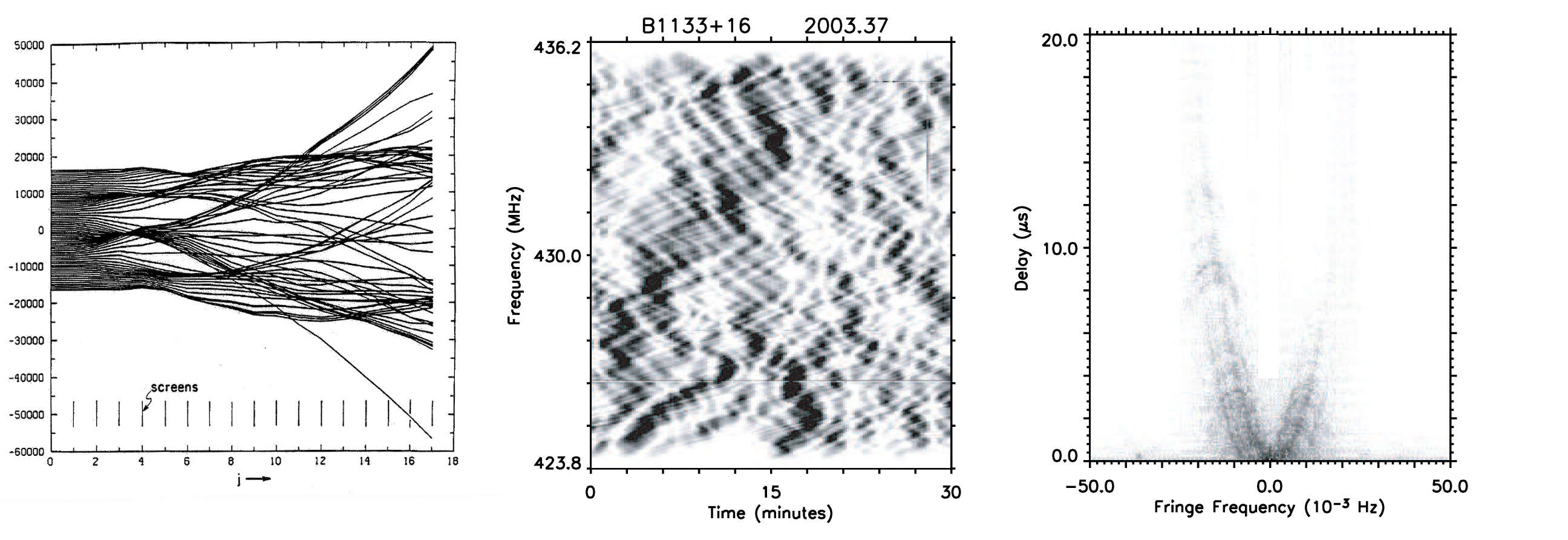}
    	\caption{ \textbf{(Left) Ray tracing by A.~Pidwerbetsky of a model with waves encountering 20 semi-random phase screens representing refracting interstellar clouds (reproduced with permission from A.~Pidwerbetsky, PhD thesis Cornell University, 1988 \cite{Pidwerbetsky1988}).
    	(Middle)  Frequency dependent signal strength as a function of time from   pulsar B0834+06, taken at  the Arecibo telescope. (Right) Double Fourier transform of the pulsar B0834+06 data.
    	Middle and right panel reproduced with permission from D.~Stinebring \cite{Stinebring} .
    	}
    	\label{fig:pidwerbetsky}}
    \end{figure*}
    A key recent innovation (Stinebring\cite{Stinebring}) led to new insight and discovery:  perform a double Fourier transform of the radio telescope time-frequency  ``dynamic spectrum'' as it is called (we would prefer ``primary spectrogram'') to create a ``secondary spectrum'' (or ``Fourier spectrogram'').   The results look nothing like the primary spectrum.  They are surprising, some are beautiful, and very informative.  They often feature one or two types of  sometimes sharp parabolic arcs, different for each pulsar, see figure~\ref{fig:pidwerbetsky}.  
    
    The clouds are modeled as thin sheets and may generate something akin to kick and drift dynamics for several successive cloud encounters. As early as 1986, Cordes and Wolzczan noted ``...multipath refractive scattering must be a common occurrence'' in pulsar microwaves arriving at earth\cite{CordesMultiple}.  This is branched flow.
    
    Indeed the pulsar data is so rich that it inspires new laboratory measurements on other systems, suggesting experiments with broadband, pulsed,  point sources and detectors with timescales shorter than those in the medium (probing turbulence this way comes to mind).

    \vskip .1in
    \noindent{\bf Semiconductor 2DEG electron flow}
    \vskip .1in
    
    We were introduced to the branched flow regime by micron scale scanning probe microscope measurements of electron flow in 2DEGS (two degree of freedom electron gasses). Bob Westervelt's lab at Harvard produced spectacular images of  of electron flow in a  2DEG\cite{Aidala2007a,march8,PhysicsTodayImaging}, including high resolution ones showing interference fringes.  The  trajectories run on a mock up  of the typical potential field seen by an electron soon revealed the origin of the branched flow seen in the experiments\cite{Topinka2001}.  Electrons are supplied to the two  dimensional layer by ionized atoms in an adjacent donor layer, but the ions induce a smooth nonuniform potential field  that randomly deflects the electrons flowing from a quantum point contact. 
    \vskip .1in
    \noindent{\bf Ocean Freak Waves}
    \vskip .1in
    Shortly after the 2DEG electron flow data was understood to be a result of branched flow, the question arose  as to other wave phenomena that might be living undetected in the branched flow regime.  The propagation of wave energy from a storm system through random refracting current eddies in the ocean comes to mind.  With one exception, the field of extreme ocean waves had made a jump from a theory of uniform sampling Gaussian random statistics pioneered by Christopher Longuette-Higgins, to important considerations of nonlinear wave interactions. The impetus for this jump was the field observation of far more frequent freak wave events than uniform Gaussian random statistics could account for.  White and Fornberg\cite{White1998}  were the exception; they attributed freak waves to the linear wave focussing due to ocean gyre lenses, using a simple incident plane wave as an example. Their studies gave a clear, early example of branched flow.  This came under heavy criticism, since no dispersion in the initial wave propagation direction was considered. Critics argued that dispersion would wash out the effects of caustics, which White and Fornberg thought to be the genesis of freak waves.  In fact, dispersion of initial wave direction (a case of fuzzy manifolds)  does not wipe out even large fluctuations in wave energy downstream. By including those fluctuations in a non-uniform Gaussian sampling scheme,  a factor of 50 more freak wave events than with uniform sampling became plausible, requiring no nonlinear wave interactions\cite{freakGeo}. 
    
    The dispersion of wave propagation directions is a classic case of a manifold of initial conditions being replaced by a fuzzy manifold.  The branching, while somewhat smoothed off by a 10\% or 20\% dispersion in initial wave direction, actually survives in quite high contrast and surprisingly develops more contrast and even finer structure downstream (figure~\ref{fig:freak}).
     \begin{figure*}
    	\includegraphics[width=5in]{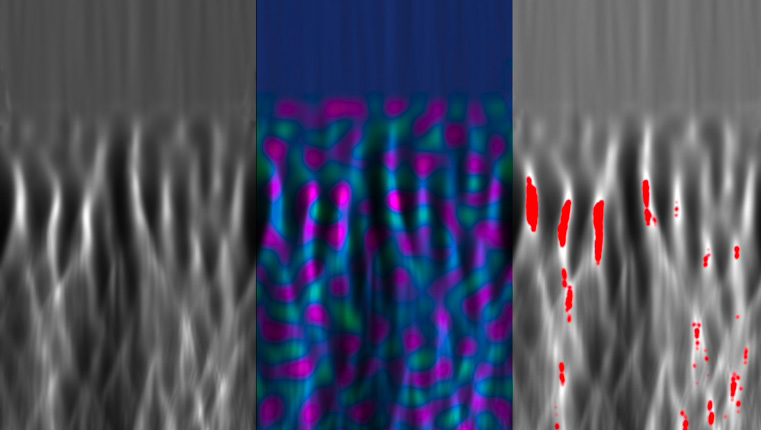}
    	\caption{ \textbf{Fuzzy manifolds and freak waves}  At the left is seen a long time average of the energy in a wave simulation using a 10 degree range of wave propagation directions,  randomly phased,  incident from the top, impinging on the random potential field shown in color in the middle panel. The red areas at the right show where large amplitude freak wave events occurred during the simulation.  The vertical height of the images is approximately 50 wavelengths of the incident waves. Freak events  happened transiently and randomly as a result of the increased wave energy in the branches and vagaries of the phases and amplitudes of the incoming wave set. Note that fine detail and large amplitude fluctuations appear even quite far from the onset of refraction.
    		\label{fig:freak}}
    \end{figure*}
    The excess  freak wave probability was accounted for without involving nonlinear effects (not denying their crucial importance in the ``end game'' of the formation of a very large wave \cite{Green2019}) by {\it nonuniform} Gaussian  sampling  over the energy density of branches of the branches surviving the fuzzy manifold. 
It is thus possible that the ignition step of freak waves is still in the linear regime.

        The connection of branched flow and ocean freak waves has inspired studies of freak wave formation in microwave cavities\cite{microwaveFreak,Barkhofen2013} and in optics for populating branches \cite{Brandstotter2019}
    \vskip .1in
    \noindent{\bf Summary}
    \vskip .1in
    
    We have mentioned some of the earliest examples we are aware of that we now call branched flow.  
    The universality of the phenomenon and huge range of applications at astronomically different scales suggests it is worthy of more effort. Branched flow lies in the interesting regime between the first focal cusps and eventual random diffusive scattering in a refractive medium with length scales over a wavelength. 
    Importantly, what can be learned about the refracting medium given knowledge of the source and the arriving signal?

    \begin{acknowledgments}
    For the authors, the branched flow regime presented itself through the remarkable experiments of Mark Topinka and Bob Westervelt of Harvard University, involving electron flow in a two degree of freedom electron gas\cite{PhysicsTodayImaging}.
    We thank Hans-J\"urgen St\"ockmann, Scott Shaw, Jiri Vanicek, Henri Degueldre,  Erik Schultheiss, Jakob Metzger, Anna Klales, Byron Drury, Robert Lin, Steve Tomsovic for valuable collaborations and discussions.  \end{acknowledgments}

    \end{document}